\documentstyle[12pt]{article}
\begin{document}
\centerline{\Large\bf Lorenz gauge quantization in }
\centerline{\Large\bf conformally flat spacetimes }
\vskip .7in
\renewcommand{\thefootnote}{\fnsymbol{footnote}}
\centerline{Jesse C. Cresswell\footnote{Present address: Department of Physics, University of Toronto, Canada M5S 1A7} and Dan N. Vollick}
\vskip .2in
\centerline{Irving K. Barber School of Arts and Sciences}
\centerline{University of British Columbia Okanagan}
\centerline{3333 University Way}
\centerline{Kelowna, B.C.}
\centerline{Canada}
\centerline{V1V 1V7}
\vskip 0.5in
\centerline{\bf\large Abstract}
\vskip 0.5in
Recently it was shown that Dirac's method of quantizing constrained dynamical systems can be used to impose the Lorenz gauge condition in a four-dimensional cosmological spacetime. In this paper we use Dirac's method to impose the Lorenz gauge condition in a general four-dimensional conformally flat spacetime and find that there is no particle production. We show that in cosmological spacetimes with dimension $D\neq4$ there will be particle production when the scale factor changes, and we calculate the particle production due to a sudden change.\newpage
\section{Introduction}
Recently it was shown \cite{Vol1} that Dirac's method of quantizing constrained dynamical systems \cite{Dir1,Dir2} can be used to impose the Lorenz gauge condition in a four-dimensional cosmological spacetime. This method was used to show that there is no particle production agreeing with earlier results by Parker \cite{Par1,Par2}.

In this paper a gauge fixed Lagrangian is introduced for the electromagnetic field in a conformally flat spacetime of arbitrary dimension. The Lorenz gauge condition is imposed as a gauge constraint and requires a secondary constraint for consistency. There are no further constraints and both constraints are first class.

In four spacetime dimensions the Hamiltonian simplifies greatly. Due to the constraints imposed on the wave function the Hamiltonian can be quantized using the flat spacetime procedure. As a result there is no particle production in four-dimensional conformally flat spacetimes. The Lagrangian introduced in this paper produces a simpler Hamiltonian and simpler constraints than the one used in Ref. \cite{Vol1} for the four-dimensional case.

We also use the gauge fixed Lagrangian to show that there will generally be particle production in $D$ dimensional cosmological spacetimes due to the changing scale factor. We then calculate the particle production that occurs for a sudden change in scale factor. The Hamiltonian is not bounded under such a transition so the wave function of the system does not remain unchanged. We show that the state vector picks up a phase factor and we calculate the resulting particle production.

\section{Hamiltonian and Constraints in $D$ Dimensions}

Consider a $D$ dimensional conformally flat spacetime with a metric
\begin{equation}
ds^2=a^2(x^{\mu})\left[-dt^2+dx_1^2+\;. . .  \;+ dx_{D-1}^2\right].
\end{equation}
The gauge fixed Lagrangian
\begin{equation}
L=-\frac{1}{4}\sqrt{g}F_{\mu\nu}\tilde{F}^{\mu\nu}-\frac{1}{2}\sqrt{g}\left(\nabla_{\mu}\tilde{A}^{\mu}\right)^2,
\end{equation}
where $\tilde{F}^{\mu\nu}=g^{\mu\alpha}g^{\nu\beta}F_{\alpha\beta}$ and $\tilde{A}^{\mu}=g^{\mu\nu}A_{\nu}$ can, after integration by parts, be written as
\begin{equation}
L=-\frac{1}{2}\sqrt{g}\left[\left(\nabla_{\mu}A_{\nu}\right)\left(\nabla^{\mu}\tilde{A}^{\nu}\right)-R_{\mu\nu}\tilde{A}^{\mu}\tilde{A}^{\nu}\right].
\end{equation}
For the above metric, this becomes, after additional integration by parts,
\begin{equation}
L=-\frac{1}{2}b\left(\partial_{\mu}A_{\nu}\right)\left(\partial^{\mu}A^{\nu}\right)
-2b(\psi_{\mu}A^{\mu})(\partial_{\nu}A^{\nu})-\frac{1}{2}b\left[4(D-3)\psi_{\mu}\psi_{\nu}-
(D-4)\partial_{\mu}\psi_{\nu}\right]A^{\mu}A^{\nu},
\end{equation}
where $A^{\mu}=\eta^{\mu\nu}A_{\nu}$, $\partial^{\mu}=\eta^{\mu\nu}\partial_{\nu}$, $b=a^{D-4}$ and
\begin{equation}
\psi_{\mu}=\frac{1}{a}\partial_{\mu}a.
\end{equation}
The Lorenz gauge condition
\begin{equation}
\nabla^{\mu}A_{\mu}=0
\end{equation}
can be written as
\begin{equation}
\partial_{\mu}A^{\mu}+(D-2)\psi_{\mu}A^{\mu}=0.
\end{equation}
The canonical momenta are
\begin{equation}
\Pi^{\mu}=b\dot{A}^{\mu}+2b\delta^{\mu}_t(\psi_{\alpha}A^{\alpha})
\end{equation}
and the Lorenz gauge condition, written in terms of the canonical momenta, is
\begin{equation}
\chi_1=\frac{1}{b}\Pi^t+\partial_kA^k+(D-4)\psi_{\mu}A^{\mu}=0.
\end{equation}
The Hamiltonian density is given by
\begin{equation}
h=\frac{1}{2b}\Pi^{\mu}\Pi_{\mu}+\frac{1}{2}b(\partial_kA_{\mu})(\partial^kA^{\mu})
-\frac{1}{2}b(D-4)(\partial_{\mu}\psi_{\nu})A^{\mu}A^{\nu}+2b(\psi_{\alpha}A^{\alpha})\chi_1.
\end{equation}
For consistency it is necessary that
\begin{equation}
\dot{\chi_1}=\{\chi_1,H\}+\frac{\partial\chi_1}{\partial t}\approx 0,
\end{equation}
where $\{\;\;\}$ denotes the Poisson bracket, $H=\int hd^{\,(D-1)}x$, and $\approx$ denotes a weak equality.
This condition gives a secondary constraint,
\begin{equation}
\chi_2=\partial_k\left(\partial^kA^t+\frac{1}{b}\Pi^k \right)+(D-4)\psi_k \left(\partial^kA^t+\frac{1}{b}\Pi^k\right)\approx 0.
\end{equation}
It is interesting to note that $\chi_2=\frac{1}{b}\partial_k(bF^{k0})$. The condition $\dot{\chi}_2\approx 0$
does not produce a new constraint, so the procedure terminates here.
The constraints $\chi_1$ and $\chi_2$ are first class since
$\{\chi_1(x),\chi_2(y)\}=0$.

\section{Quantization in Four Dimensions}
In four spacetime dimensions, the Hamiltonian is given by
\begin{equation}
H=\int\left[\frac{1}{2}\Pi^{\mu}\Pi_{\mu}+\frac{1}{2}(\partial_kA_{\mu})(\partial^kA^{\mu})
+2(\psi_{\alpha}A^{\alpha})\chi_1\right]d^3x,
\label{Ham}
\end{equation}
while the constraints are given by
\begin{equation}
\chi_1=\Pi^t+\partial_kA^k
\end{equation}
and
\begin{equation}
\chi_2=\partial_k\left(\partial^kA^t+\Pi^k \right).
\end{equation}
These are identical to the expressions in flat spacetime, except for the last term in the Hamiltonian involving $\chi_1$.

To quantize the theory we follow the procedure developed by Dirac \cite{Dir1,Dir2}.
In the Schr\"{o}dinger picture, the dynamical variables $A_{\mu}$ and $\Pi^{\mu}$ become time independent operators satisfying
\begin{equation}
[A_{\mu}(\vec{x}),A_{\nu}(\vec{y})]=[\Pi^{\mu}(\vec{x}),\Pi^{\nu}(\vec{y})]=0
\end{equation}
and
\begin{equation}
[A_{\mu}(\vec{x}),\Pi^{\nu}(\vec{y})]=i\,\delta^{\nu}_{\mu}\,\delta^3(\vec{x},\vec{y}),
\end{equation}
where $[\;\;]$ denotes the commutator and we have set $\hbar=1$. Note that there is an ambiguity in the ordering of the operators in the last term in the Hamiltonian since $\chi_1$ contains $\Pi^t$ and $\psi^{\alpha}A_{\alpha}$ contains $A_t$. We have chosen the ordering so that the Hamiltonian is given by (\ref{Ham}).

A state vector is introduced that
satisfies the Schr\"{o}dinger equation
\begin{equation}
i\frac{d}{dt}|\Psi>=H|\Psi>.
\end{equation}
The constraints are imposed on the wave function as follows:
\begin{equation}
\chi_1\,|\Psi>=0\;\;\;\;\;\; \mathrm{and}\;\;\;\;\;\;\chi_2\,|\Psi>=0.
\end{equation}
The last term in the Hamiltonian will therefore not affect the equations of motion, and the theory can be quantized by following the flat spacetime procedure.

This generalizes the results of Ref. \cite{Vol1} from a four-dimensional cosmological spacetime to a general four-dimensional conformally flat spacetime. Thus, there is no particle production in four-dimensional conformally flat spacetimes, as expected based on Parker's calculation for a massless conformally coupled scalar field \cite{Par3}. The Lagrangian used in this paper differs from the one used in Ref. \cite{Vol1} by a total derivative and gives a simpler $H$, $\chi_1$ and $\chi_2$.

\section{Quantization in $D$-dimensional Cosmological Spacetimes}
To examine particle production in spacetimes with $D\neq 4$ we consider the case in which
$a$ depends only on $t$. In this case $\psi_k=0$, and we write $\psi_t=\psi$.

The Hamiltonian is
\begin{equation}
H=\frac{1}{2}\int\left[\frac{1}{b}\Pi^{\mu}\Pi_{\mu}+b(\partial_kA_{\mu})(\partial^kA^{\mu})
-4b\psi A_t\chi_1-b(D-4)\dot{\psi}A_t^2
\right]d^{\,(D-1)}x,
\end{equation}
while the constraints are
\begin{equation}
\chi_1=\frac{1}{b}\Pi^t+\partial_kA^k-(D-4)\psi A_t
\end{equation}
and
\begin{equation}
\chi_2=\partial_k\left(\partial^kA^t+\frac{1}{b}\Pi^k\right).
\end{equation}

For a consistent quantum theory we require that $[\chi_1,\chi_2]=
\alpha\chi_1+\beta\chi_2$, where $\alpha$ and $\beta$ are operators that
appear to the left of the constraints. This is satisfied since $[\chi_1,\chi_2]=0$.

To preserve the constraints under time evolution it is necessary that
\begin{equation}
\frac{\partial\chi_k}{\partial t}-i\left[\chi_k,H\right]\approx 0,
\end{equation}
where, in the quantum theory, $A\approx 0$ implies that
$A|\Psi>=0$. It is easy to show that the constraints are preserved, as they are in the classical case.

The constraint $\chi_2$ can be simplified. The term $\partial_m\Pi^m$ involves
only the longitudinal part of $\Pi^m$, and this longitudinal part can be written
as the gradient of a scalar $U$. Thus, $\partial_m\Pi^m=\nabla^2U$. The constraint
$\chi_2$ can therefore be written as
\begin{equation}
\chi_2=\nabla^2\left(A^t+\frac{1}{b}U\right).
\end{equation}
Now, $\nabla^2(A^t+\frac{1}{b}U)\approx 0$ over all space has the unique solution $A^t+\frac{1}{b}U
\approx 0$ if the fields vanish at infinity. Since we are quantizing the electromagnetic field on a fixed background spacetime we are free to make this assumption. 

The Hamiltonian can be decomposed into transverse and longitudinal/timelike
parts:
\begin{equation}
H_T=\frac{1}{2}\int\left[\frac{1}{b}\Pi^m_{(T)}\Pi^{(T)}_m+b\left(\partial_sA^{(T)}_m\right)
\left(\partial^sA^m_{(T)}\right)\right]d^{\,(D-1)}x,
\label{HT}
\end{equation}
\begin{equation}
H_{L}^{(1)}=\frac{b}{2}\int\left[\partial_r(\frac{1}{b}U-A^t)\partial^r(\frac{1}{b}U+A^t)+
(\partial_mA^m
-\frac{1}{b}\Pi^t-D\psi A_t)\chi_1\right]d^{\,(D-1)}x,
\end{equation}
and
\begin{equation}
H_{L}^{(2)}=\frac{1}{2}(D-4)b\int\{2\psi A_t(\partial_kA^k)-[(D-4)\psi^2+\dot{\psi}]A_t^2\}d^{\,(D-1)}x.
\end{equation}
Note that $H_{L}^{(1)}\approx 0$, so that $H\approx H_T+H_L^{(2)}$.

To set up a Fock space representation in the Minkowski in and out regions, where $a$ is constant, the operators
\begin{equation}
a^{(\lambda)}_{\vec{k}}=\int e^{-i\vec{k}\cdot\vec{x}}\left[k \sqrt{b}\epsilon_{\vec{k}\mu}^{(\lambda)}A^{\mu}(\vec{x})+
\frac{i}{\sqrt{b}}\epsilon_{\vec{k}\mu}^{(\lambda)}\Pi^{\mu}(\vec{x})\right] d^{\,(D-1)}x
\end{equation}
and
\begin{equation}
a^{\dag\,(\lambda)}_{\vec{k}}=\int e^{i\vec{k}\cdot\vec{x}}\left[k \sqrt{b}\epsilon_{\vec{k}\mu}^{(\lambda)}A^{\mu}(\vec{x})-
\frac{i}{\sqrt{b}}\epsilon_{\vec{k}\mu}^{(\lambda)}\Pi^{\mu}(\vec{x})\right]d^{\,(D-1)}x
\end{equation}
can be introduced. Here $k=|\vec{k}|$ and $\epsilon_{\vec{k}\mu}^{(\lambda)}$ are the standard (real) Minkowski polarization vectors.
The factors of $b$ in the above expressions can be deduced by computing these
operators using the standard approach in the Heisenberg picture and then transforming them into the Schr\"{o}dinger picture.
These are also the unique factors of
$b$ that give the standard commutation relations
\begin{equation}
\left[a^{(\lambda)}_{\vec{k}}\,,\,a^{(\lambda')}_{\vec{k}'}\right]=\left[a^{\dag\,(\lambda)}_{\vec{k}}\,,
\,a^{\dag\,(\lambda')}_{\vec{k}'}\right]=0,
\end{equation}
\begin{equation}
\left[a^{(\lambda)}_{\vec{k}}\,,\,a^{\dag\,(\lambda')}_{\vec{k}'}\right]=(2\pi)^{(D-1)}(2k)\eta^{\lambda\lambda'}
\delta^{(D-1)}(\vec{k}-\vec{k}')
\end{equation}
and normal ordered Hamiltonian
\begin{equation}
{:H_{T}:} =\frac{1}{2(2\pi)^{(D-1)}}\sum_{\lambda=1}^{D-2}\int a^{\dag}_{(\lambda)\vec{k}}a^{(\lambda)}_{\vec{k}}
\, d^{\,(D-1)}k.
\end{equation}
Note that the polarizations corresponding to $\lambda=1...(D-2)$ are transverse polarizations.

A vacuum state $|0>$ can be introduced that satisfies
\begin{equation}
\left[\frac{1}{b}\Pi^t+\partial_mA^m\right]|0>=0,\;\;\;\;\;\;\;\;\;\;
\left[\frac{1}{b}\partial_m\Pi^m+\nabla^2A^t\right]|0>=0,
\label{vac1}
\end{equation}
and
\begin{equation}
a^{(\lambda)}_{\vec{k}}\,|0>=0 ,\;\;\;\;\;\;\;\;\;\;\;\;\;\; \lambda=1...(D-2).
\label{vac2}
\end{equation}
The operators $a^{(\lambda)}_{\vec{k}}$ act as annihilation operators, and the
operators $a^{\dag (\lambda)}_{\vec{k}}$ act as creation operators. Note that
$|0_{\mathrm{in}}>$ will not be the same as $|0_{\mathrm{out}}>$ if $b_{\mathrm{out}}\neq b_{\mathrm{in}}$.
There will therefore be particle production unless the in-vacuum state happens to evolve into the out-vacuum state.

As an explicit example of particle production, consider the case of a ``sudden" change from a
Minkowski space with $b_{\mathrm{in}}$ to one with $b_{\mathrm{out}}$.
The sudden approximation cannot be used because
the Hamiltonian contains terms involving $\psi$ and $\dot{\psi}$ which do not remain bounded
as the time interval over which the change takes place goes to zero.
The behavior of the state vector can be determined by introducing the ket $|\Psi_T>$,
\begin{equation}
|\Psi_T>=\exp\left\{ib\int\left[A_t(\partial_{k}A^k)-\frac{1}{2}(D-4)\psi A_t^2\right]d^{(D-1)}x\right\}|\Psi>,
\end{equation}
which satisfies the equation of motion
\begin{equation}
i\frac{d}{dt}|\Psi_T>=H_T|\Psi_T>
\end{equation}
and the constraints
\begin{equation}
\Pi^t|\Psi_T>=0\;\;\;\;\;\;\;\;\;\;\;\;\; \mathrm{and} \;\;\;\;\;\;\;\;\;\;\;\;\; {\partial_{\mathnormal{k}}\Pi^{\mathnormal{k}}|\Psi_{\mathnormal{T}}>=0}.
\end{equation}
Since $H_T$ (\ref{HT}) remains bounded during the transition, the sudden approximation
can be used on the evolution of $|\Psi_T>$. This means that $|\Psi_T>$ does not change and that $|\Psi>$ picks up a phase factor during the transition. Thus, if the initial state is the in-vacuum, then the state in the out region, just after the transition, will be the in-vacuum with a phase factor. Since the phase factor commutes with the transverse $a^{(\lambda)}_{(\mathrm{in})\vec{k}}$ the state of the system will still be in the in-vacuum state.

The Bogolubov
transformation between the in and out operators is given by
\begin{equation}
a^{(\lambda)}_{(\mathrm{out})\vec{k}}=\left[\frac{b_{\mathrm{in}}+b_{\mathrm{out}}}{2\sqrt{b_{\mathrm{in}}b_{\mathrm{out}}}}\right]a^{(\lambda)}_{(\mathrm{in})\vec{k}}
\pm\left[\frac{b_{\mathrm{out}}-b_{\mathrm{in}}}{2\sqrt{b_{\mathrm{in}}b_{\mathrm{out}}}}\right]a^{\dag(\lambda)}_{(\mathrm{in})-\vec{k}}
\end{equation}
where we have taken $\epsilon_{(-\vec{k})\mu}^{(\lambda)}=\pm \epsilon_{\vec{k}\mu}^{(\lambda)}$.
The expectation value of $N_{(\mathrm{out})\vec{k}}^{(\lambda)}=a_{(\mathrm{out})\vec{k}}^{\dag (\lambda)}
a_{(\mathrm{out})\vec{k}}^{(\lambda)}$ in the in-vacuum state is
\begin{equation}
<0_{\mathrm{in}}|N^{(\lambda)}_{(\mathrm{out})\vec{k}}|0_{\mathrm{in}}>=\frac{(b_{\mathrm{out}}-b_{\mathrm{in}})^2}{4b_{\mathrm{in}}b_{\mathrm{out}}}.
\end{equation}
There will therefore be particles produced by the sudden change in the scale factor when $D\neq4$.
\section{Conclusion}
In this paper we used Dirac's method of quantizing constrained dynamical systems to generalize the results of Ref. \cite{Vol1}. We found that in four-dimensional conformally flat spacetimes the Hamiltonian and constraints have the same form as in flat space but for an extra term in the Hamiltonian. Due to the constraints on the system, the extra term has no effect on the equations of motion, so there is no particle production in agreement with Ref. \cite{Par3}.

We also considered cosmological spacetimes with $D\neq4$ and found that there is particle production unless the in-vacuum state happens to evolve into the out-vacuum state. For a spacetime that undergoes a sudden change in scale factor the wave function of the system picks up a phase factor because the Hamiltonian does not remain bounded. Under the sudden change, we found that if the initial state is the in-vacuum the final state will be the in-vacuum with a phase factor. A Bogolubov transformation between the in and out creation and annihilation operators showed that the expectation value of the out-number operator in the in-vacuum state is ${(b_{\mathrm{out}}-b_{\mathrm{in}})^2}/{4b_{\mathrm{in}} b_{\mathrm{out}}}$ where $b=a^{D-4}$ and $a(t)$ is the scale factor of the spacetime.
\section*{Acknowledgements}
This research was supported by the Natural Sciences and Engineering Research
Council of Canada.


\begin{thebibliography}{99}

\bibitem{Vol1}
D. N. Vollick, Phys. Rev. D \textbf{86}, 084057 (2012).
\bibitem{Dir1}
P.A.M. Dirac, \emph{The Principles of Quantum Mechanics}, 4th ed. (Oxford
University Press, New York, 1958),  Chap. VII.
\bibitem{Dir2}
P.A.M. Dirac, \emph{Lectures on Quantum Mechanics} (Dover, New York, 1964).
\bibitem{Par1}
L. Parker, Phys. Rev. Lett. \textbf{21}, 562 (1968).
\bibitem{Par2}
L. Parker, Phys. Rev. \textbf{183}, 1057 (1969).
\bibitem{Par3}
L. Parker, Phys. Rev. D \textbf{7}, 976 (1973).
\end{thebibliography}
\end{document}